\documentclass[conference]{IEEEtran}
\IEEEoverridecommandlockouts
\usepackage{cite}
\usepackage{amsmath,amssymb,amsfonts}
\usepackage{algorithmic}
\usepackage{graphicx}
\usepackage{textcomp}
\usepackage{xcolor}
\usepackage{hyperref}
\usepackage{comment}

\def\BibTeX{{\rm B\kern-.05em{\sc i\kern-.025em b}\kern-.08em
    T\kern-.1667em\lower.7ex\hbox{E}\kern-.125emX}}
\begin{document}

\title{Diffusion Models are Robust Pretrainers}

\author{\IEEEauthorblockN{Mika Yagoda}
\IEEEauthorblockA{\textit{Electrical Engineering Department} \\
\textit{Tel Aviv University}\\
Tel Aviv, Israel \\
yagodamika@mail.tau.ac.il}
\and
\IEEEauthorblockN{Shady Abu-Hussein}
\IEEEauthorblockA{\textit{Electrical Engineering Department} \\
\textit{Tel Aviv University}\\
Tel Aviv, Israel \\
shady.abh@gmail.com}
\and
\IEEEauthorblockN{Raja Giryes}
\IEEEauthorblockA{\textit{Electrical Engineering Department} \\
\textit{Tel Aviv University}\\
Tel Aviv, Israel \\
raja@tauex.tau.ac.il }

}

\maketitle

\begin{abstract}
Diffusion models have gained significant attention for high-fidelity image generation. Our work investigates the potential of exploiting diffusion models for adversarial robustness in image classification and object detection. Adversarial attacks challenge standard models in these tasks by perturbing inputs to force incorrect predictions. To address this issue, many approaches use training schemes for forcing the robustness of the models, which increase training costs.
In this work, we study models built on top of off-the-shelf diffusion models and demonstrate their practical significance: 
they provide a low-cost path to robust representations, allowing lightweight heads to be trained on frozen features without full adversarial training. Our empirical evaluations on ImageNet, CIFAR-10, and PASCAL VOC show that diffusion-based classifiers and detectors achieve meaningful adversarial robustness with minimal compute. 
While clean and adversarial accuracies remain below state-of-the-art adversarially trained CNNs or ViTs, diffusion pretraining offers a favorable tradeoff between efficiency and robustness. 
This work opens a promising avenue for integrating diffusion models into resource-constrained robust deployments.

Code is available at \href{https://github.com/yagodamika/Diffusion-Models-are-Robust-Pretrainers}{\textcolor{purple}{https://github.com/yagodamika/Diffusion-Models-are-Robust-Pretrainers}}
\end{abstract}


\section{Introduction}

Diffusion models have recently achieved state-of-the-art performance in high-fidelity image generation \cite{ho2020denoising, abu2023udpm, karras2022elucidating, song2020score}, outperforming prior generative models \cite{dhariwal2021diffusion} with more stable training and scalability to higher resolutions and diverse datasets \cite{rombach2022high}. Diffusion models define a Markovian process that starts with clean images on the one end and pure noise of a known distribution on the other end. The forward process progressively adds noise to images, while the reverse process trains a neural network to denoise, generating images from noise to the clean domain.

Adversarial robustness refers to the ability of neural networks to resist adversarial attacks. These attacks involve manipulating input data in ways that are imperceptible to humans but cause models to make err. For example, in image classification, an attacker perturbs the input image with small, carefully crafted noise that is indistinguishable to humans but leads the model to make an incorrect class prediction \cite{goodfellow2014explaining, madry2017towards, croce2020reliable}.

Multiple approaches have been proposed for robustifying deep neural networks against adversarial attacks. One direct approach integrates the adversarial examples within the training of the model, often referred to as "adversarial training'" \cite{mkadry2017towards}. Other approaches suggest to use some sort of regularization \cite{li2021towards, amini2020towards, esmaeilpour2020detection, serrurier2021achieving, liu2021training, zhang2020interpreting}, or employ a specific network architecture \cite{yu2021defending, li2021wavecnet}.

Self-supervised learning (SSL) pretrains models on large unlabeled datasets \cite{donahue2019large, li2023mage, he2022masked}, producing rich representations for downstream tasks. It has been demonstrated \cite{chen2020adversarial} that adversarial training can be incorporated into the unsupervised training stage, resulting in significant performance improvements compared to conventional end-to-end adversarial training baselines \cite{chen2020adversarial}. Recently, pretrained diffusion models have also been successfully transferred to downstream tasks, surpassing other generative pretrainers \cite{mukhopadhyay2023diffusion}, but their robustness remains unexplored. \cite{chen2023robust} shows that diffusion classifiers exploiting conditional-unconditional score differences exhibit inherent robustness.  Yet, they rely on labeled training and in this work we focus on unconditional diffusion models. 

We investigate whether diffusion models pretrained without labels provide robust features against adversarial attacks. We train lightweight classification heads on top of frozen features from off-the-shelf unconditional diffusion models and extend this method to object detection. Our results show that diffusion-based features offer robustness for both tasks.

The main benefit of our approach is efficiency: it avoids costly adversarial training and requires only a lightweight head on frozen features. This makes it especially attractive for low-compute deployments or scenarios with limited labeled data. While diffusion features provide robustness “for free,” clean and adversarial accuracies remain below those of state-of-the-art adversarially trained CNNs or ViTs, positioning our method as a computationally efficient alternative that offers robustness without additional training overhead. 

Experiments are conducted on CIFAR-10 \cite{cifar10} and ImageNet \cite{deng2009imagenet} for classification, and PASCAL VOC for object detection. We study the effect of layer (block) and diffusion timestep choices on robustness, revealing that timestep choice has a greater impact on robustness than layer selection.

\begin{figure}[t]
\vspace{-0.76in}
\centerline{\includegraphics[width=1.8\columnwidth]{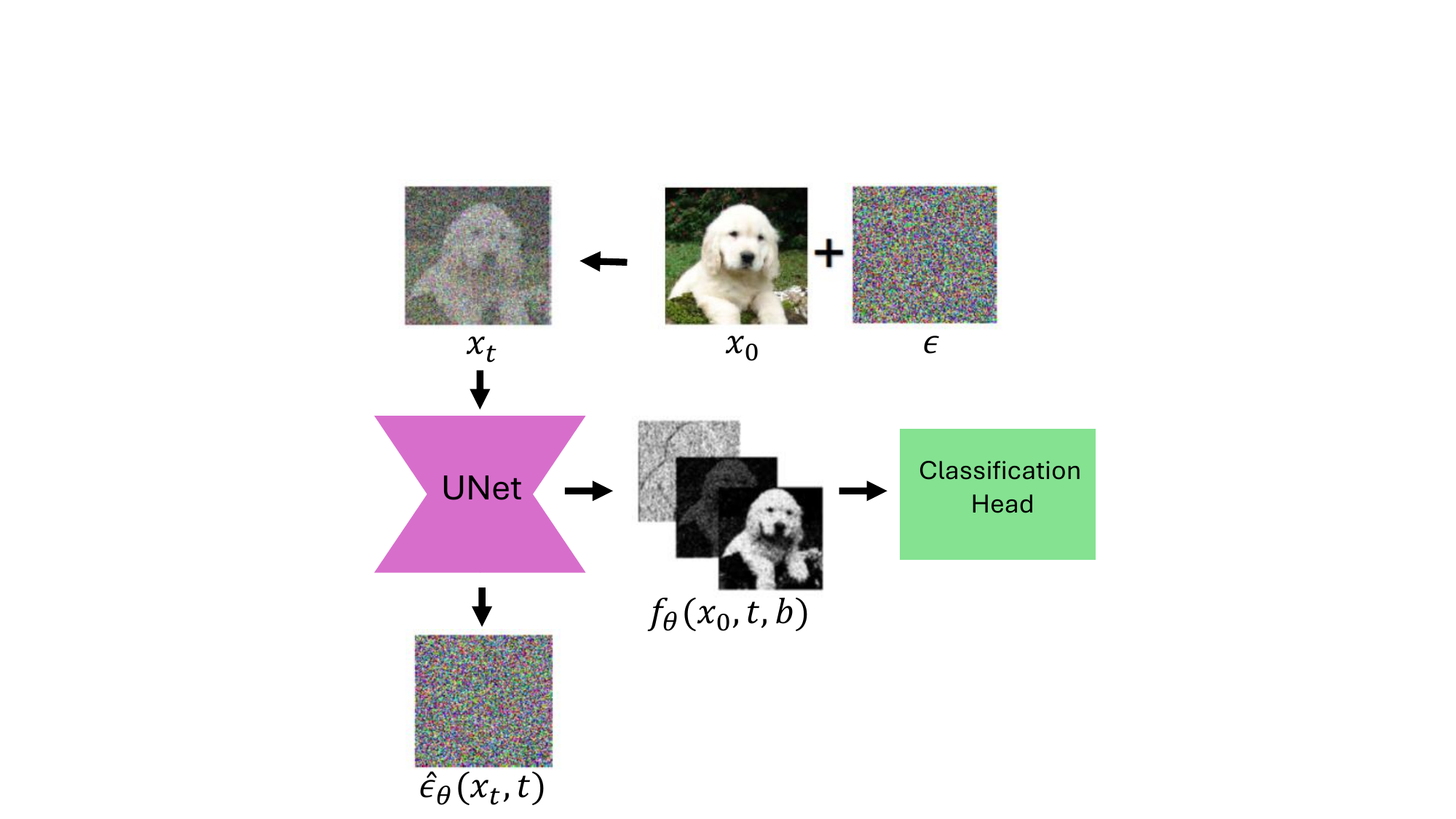}}
\caption{An overview of the used diffusion-based classification method. 
We examine the robustness of these models with respect to the U-Net block number and the diffusion noise time step. We use lightweight architectures for feature classification, including linear and attention-based heads. \label{fig1}}
\end{figure}

\section{Background}

\subsection{Adversarial Attacks}
We examine a deep classifier model $ f_\gamma : \mathbb{R}^N \rightarrow \mathbb{R}^C $, where $ N $, $C$, and $\gamma$ denote the input image dimension, the number of classes, and the classifier parameters respectively. Adversarial examples are inputs intentionally crafted by an attacker to induce incorrect predictions by $ f_\gamma $. 
These examples are generated by adding a small perturbation to the input image, such that is indistinguishable perceptually, yet leads to totally different class prediction. 
Formally, given an input $ x $, its true label $ y $, and a threat set $\Delta = \{ \delta : \|\delta\|_{n\in \{2, \infty\}} \leq \epsilon \}$, an adversarial example $\hat{x}$ is given by
\begin{equation*}
\hat{x} = x + \delta \quad \text{where } \delta \in \Delta \text{ and } f_\gamma(\hat{x}) \neq y,
\end{equation*}

The process of generating such examples is referred to as an adversarial attack and can be categorized into untargeted or targeted attacks. Untargeted attacks aim to generate $ \hat{x} $ leading to misclassifications without a specific target class. While targeted attacks aim to create $ \hat{x} $ inducing the classifier to predict $ \hat{x} $ as some traget class $ \hat{y} \neq y$. There are various methods for creating adversarial examples. One known attack is the Projected Gradient Descent (PGD) attack \cite{mkadry2017towards}, outlined by the following iterative scheme
\begin{equation*}
\text{Repeat $n$ times: } \delta = \Pi_{\epsilon} \left( \delta + \alpha \nabla_{\delta} L(f_\gamma(x + \delta), \hat{y}) \right),
\end{equation*}
where $ \Pi_{\epsilon} $ denotes the projection operator onto $ \Delta $, $ \alpha $ is the step size, $n$ is the number of steps, and $ L(\cdot) $ denotes the cross-entropy classification loss. 

A popular approach to improve robustness to these attacks is using adversarial training. Yet, it is costly and increases the training time. Thus, we focus on methods that do not use it in the fine-tuning stage and show that our proposed approach can lead to good robustness even without it. 

\subsection{Diffusion Models}
Diffusion models define a forward noising process, which involves progressively adding Gaussian noise to an image $x_0$ sampled from the data distribution $q(x_0)$. This results in a fully noised image $x_T$ after $ T $ steps. The forward process is structured as a Markov chain with latent variables $ x_1, x_2, \ldots, x_{T-1}, x_T $, where each $ x_t $ denotes an image affected by an increasing noise level. Formally, the forward diffusion process can be expressed as:
\begin{equation*}
    q(x_1, \ldots, x_T | x_0) := \prod_{t=1}^{T} q(x_t | x_{t-1}),    
\end{equation*}
where 
\begin{equation*}
q(x_t | x_{t-1}) := \mathcal{N} \left( x_t; \sqrt{1 - \beta_t} x_{t-1}, \beta_t I \right).
\end{equation*}
and $ \{\beta_t\}_{t=1}^{T} $ controls the noise variance scheduler. By defining $ \alpha_t := 1 - \beta_t $ and $ \bar{\alpha}_t := \prod_{i=0}^{t} \alpha_i $, one can directly sample a noised image $ x_t $ at diffusion step $ t $ from the original image $ x_0 $ according to the parameterization:
\begin{equation}
x_t = \sqrt{\bar{\alpha}_t} x_0 + \sqrt{1 - \bar{\alpha}_t} \epsilon, \quad \epsilon \sim \mathcal{N}(0, I).
\label{eq:xt_given_x0}
\end{equation}

The reverse diffusion process seeks to invert the forward process and sample from the posterior distribution $q(x_{t-1} | x_t)$. It is performed by denoising $x_t$ and then adding a noise perturbation according to the noise scheduler at step $t-1$. 
By starting from $x_T$ and progressively running the backward process until reaching $x_0$, one may obtain a clean image.
This allows sampling from the original data distribution $q(x_0)$. The denoising step is approximated using a neural network with parameters $\Theta$, denoted by $ \epsilon_\Theta$, trained to predict the score $\epsilon$. 

\subsection{Diffussion Models are Robust Pretrainers}
We use the framework suggested in \cite{mukhopadhyay2023diffusion} for utilizing diffusion models for classification, as depicted in Figure \ref{fig1}.
Given an input image $x_0$, we apply the forward diffusion process to obtain a partially noised version $x_t$ at timestep $t$. To do that we fix the timestep t to a desired value and then apply the forward diffusion model using the parameterization in Equation \eqref{eq:xt_given_x0} to obtain $x_t$. The noised image $x_t$ is then passed through the frozen UNet backbone of the pretrained diffusion model $\epsilon_\Theta(x_t, t)$.

The U-Net consists of a sequence of encoder and decoder blocks operating at multiple spatial resolutions. From this network, we extract hidden feature maps $g_\theta (x_t,t, b)$ from the output of block $b$. These feature maps are then reduced to a fixed-size representation using adaptive average pooling followed by flattening, producing a 1D feature vector.

On top of this vector, we train a lightweight classification head $h_\omega$, with parameters $\omega$. We experiment with two head architectures: 1) a single linear layer and 2) a single layer self-attention module. The diffusion model parameters $\Theta$ remain frozen during training, and only the classification head parameters $\omega$ are updated.  

We show that classifiers built in such manner possess greater adversarial robustness comparing to other methods including robust pretraining methods and another generative model-based method. To evaluate the robustness of the models we use robust accuracy, the classification accuracy on an attacked test dataset. 

We explore a similar approach for object detection. We used an object detection head and fed it with feature maps extracted from a frozen diffusion backbone. In a similar way to the classification approach, we extract from the diffusion model the feature maps $g_\theta$(xt, t, b) for time step t  and block b. We pass these feature maps through simple adaptation layers - 1x1 convolutional layers or 1x1 convolutions followed by multihead attention layers - and pass their outputs into the detection head. 
During training we trained both the detection head and the adaptation layers while keeping the diffusion model parameters fixed.
To evaluate the models' robustness, we subjected them to multiple adversarial attacks.

\begin{figure}[t]
\centerline{\includegraphics[width=\linewidth]{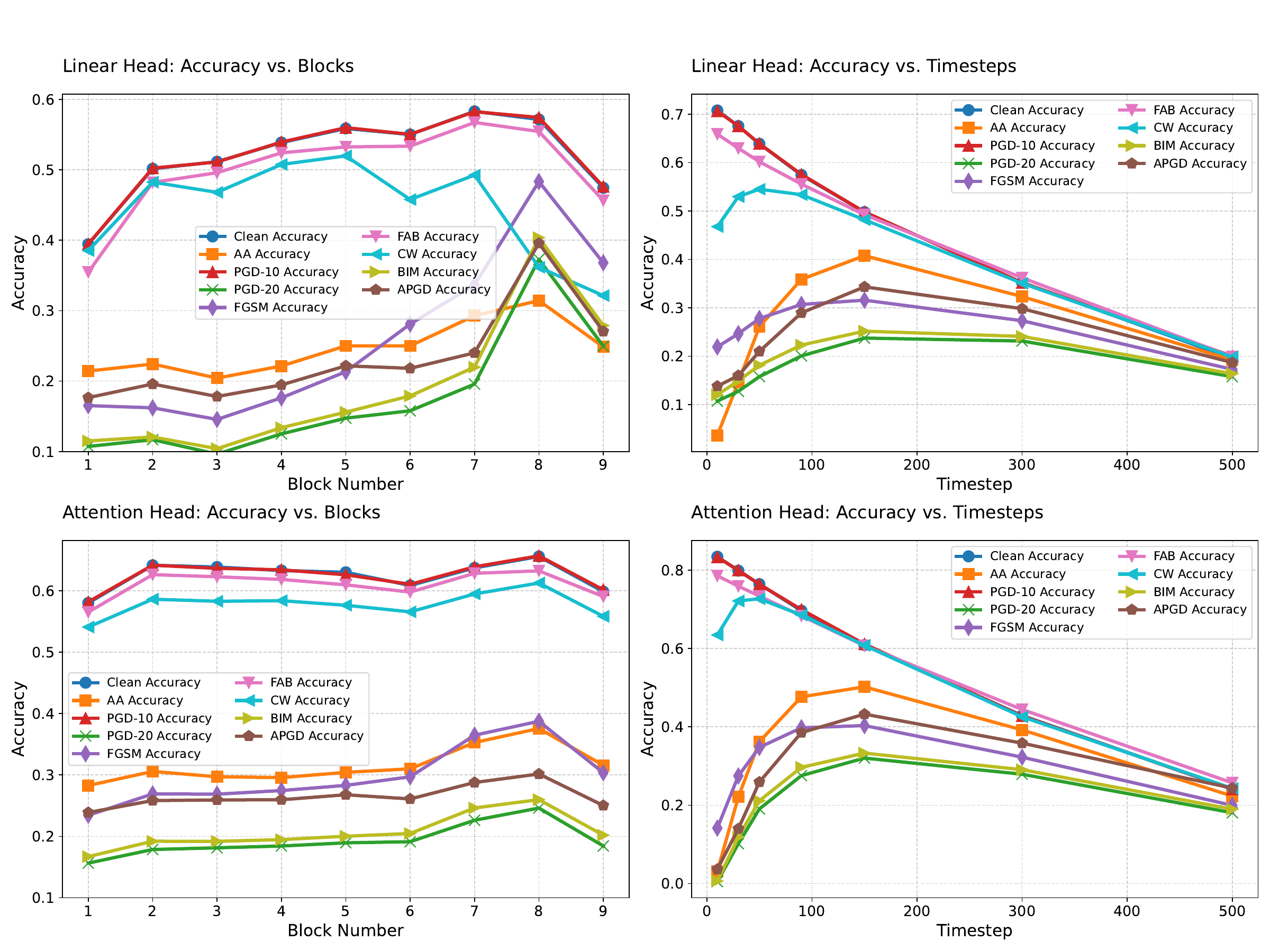}}
\caption{Ablations on CIFAR-10 with varying block numbers and time steps, for a linear and an attention classification heads on frozen features. The accuracies are averaged over timesteps or block numbers. \label{fig2}}
\end{figure}

\section{EXPERIMENTS}
We begin by providing details of the experiments setup, and then we present our main results, where we present our robustness evaluation of diffusion pretrainers and compare them to those of other models. Finally, we show ablation studies demonstrating the diffusion time step choice and the network block choice effect on the robustness of the classifier.

\subsection{Experimental Setup}
\label{sec:exp_setup}
For classification, we test the method on both ImageNet and CIFAR-10. For ImageNet we use the unconditional diffusion classification model proposed by  \cite{mukhopadhyay2023diffusion}, which uses the unconditional ADM model from \cite{dhariwal2021diffusion}. For CIFAR-10 we use the unconditional DDPM model \cite{ho2020denoising}.

We examine the robustness of two types of classification heads: 1) a single linear layer. 2) a single attention layer. We keep the diffusion model weights frozen and only train the head to predict the target class by minimizing the traditional cross-entropy loss. For CIFAR10 we use the SGD optimizer with learning rate 1e-2 and batch size set to 32. We train for 20 epochs with a learning rate decay factor of factor 0.1, decayed every 7 epochs. 
We use an adaptive average pooling to reduce the spatial dimensions of the features. While for ImageNet we use the pretrained classification models from \cite{mukhopadhyay2023diffusion}.

For object detection, we test our method on the PASCAL VOC dataset. We use the unconditional diffusion classification model proposed by  \cite{mukhopadhyay2023diffusion}, which uses the unconditional ADM model from \cite{dhariwal2021diffusion}. We use two types of layers: 1) 1x1 convolutions 2) 1x1 convolutions followed by multihead attention layers. 
We extract several feature maps from the diffusion model and insert them to these layers, and the output maps are inserted to a detection head. For the detection head we used the detection headers from the RobustDet model \cite{dong2022adversariallyawarerobustobjectdetector}.
We selected block numbers whose feature map dimensions matched the input requirements of the detection headers and demonstrated strong robustness in our classification experiments, specifically, blocks 28, 25, and 24. For timesteps, we also chose those that exhibited robust classification, specifically, t = 60, 90, and 120.

\begin{table}[t]
\caption{ImageNet Classification Results of Diffusion-Based Models}
\centering
\resizebox{\linewidth}{!}{
    \begin{tabular}{|c|c|c|c|c|c|}
    \hline
    \textbf{Head Type} & \textbf{Block \#} & \textbf{Timestep} & \textbf{Accuracy [\%]} & \textbf{PGD-10 Accuracy [\%]}  \\ \hline
    Linear    & 24 & 90  & 61.9 & 46.3  \\ \hline
    Attention & 24 & 150 & 74.3 & 39.0  \\ \hline
    \end{tabular}
}
\label{tab:imagenet_results}
\end{table}

\subsection{Main Results}
We first report the baseline clean accuracy of the models, i.e., their performance on unperturbed inputs. To assess adversarial robustness, we evaluate the models under multiple attacks, including FGSM \cite{goodfellow2014explaining}, BIM \cite{kurakin2018adversarial}, CW \cite{carlini2017towards}, FAB \cite{croce2020minimally}, APGD \cite{croce2020reliable}, AutoAttack (AA) \cite{croce2020reliable}, and PGD \cite{mkadry2017towards}. 

We compare our performance to other CIFAR-10 pretrained classifiers in Table \ref{tab:CIFAR-10_Accuracies}. Specifically, we compare to a classifier based on the FlowGMM generative model \cite{izmailov2020semi},  which consists of a linear layer trained on top of the latent features of FlowGMM that remain frozen during the training, similar to our setup. In addition to FlowGMM, we compare to models from \cite{chen2020adversarial} that were pretrained in a self-supervised manner combined with adversarial training. 
Their training included self-supervised pretraining using an adversarial loss, resulting in a mapping from an input sample to an embedding space. After pretraining, a supervised finetuning stage is performed, in which representations learned in the pretraining stage are mapped to the label space. We use the P3-F1 setup mentioned in \cite{chen2020adversarial}, where the pretraining stage includes adversarial training, while the finetuning stage includes only standard training.
We examine models that were pretrained on the following self-supervised pretraining tasks: Selfie \cite{trinh2019self}, Rotation \cite{gidaris2018unsupervised} and Jigsaw \cite{noroozi2016unsupervised,carlucci2019domain}. 
For consistency, we compare the performance of the diffusion classifiers to these models using the same PGD-20 configuration used in  \cite{chen2020adversarial}.

Table \ref{tab:CIFAR-10_Accuracies} shows that diffusion-pretrained models achieve clean performance comparable to other pretraining-based approaches, while providing a clear advantage in robustness. Importantly, our models do not employ any adversarial training. As expected, they do not yet reach the performance of state-of-the-art adversarial robustness methods, which are explicitly optimized for robustness, such as \cite{bartoldson2024adversarial},\cite{amini2024meansparse} . 

In Table~\ref{tab:CIFAR10_Attacks_Top3} we compare our method against the state-of-the-art on CIFAR-10, including a version of Google ViT-B/16 pre-trained on ImageNet-21k and finetuned on the CIFAR-10 dataset, taken from \href{https://huggingface.co/aaraki/vit-base-patch16-224-in21k-finetuned-cifar10}{Huggingface}. The ViT-B/16 baseline achieves the highest clean accuracy (97.88\%) but collapses under most attacks, dropping near zero. In contrast, our diffusion-based heads trade some clean accuracy (\(71{-}82\%\)) for markedly stronger robustness. For example, the attention head with \(b{=}8, t{=}90\) reaches 56\% under AutoAttack, showing a more balanced clean-robustness tradeoff.

For object detection, considering that the object detector has two tasks of classification and localization, we used PGD to attack the classification (CLS attack) and localization (LOC attack). We also test the robustness under the CWA attack \cite{chen2021classawarerobustadversarialtraining}. 
We used the same attack configurations used in the RobustDet paper \cite{dong2022adversariallyawarerobustobjectdetector}.
Table \ref{tab:detection_results} demonstrates that our strategy is also robust in the object detection scenario.

\begin{table}[t]
\caption{CIFAR-10 Classification Results. The best performing model is marked in bold and the second best in blue.}
\centering
\resizebox{\linewidth}{!}{
    \begin{tabular}{|c|c|c|c|c|}
    \hline
        \textbf{Model} & \textbf{Clean Accuracy [\%]} & \textbf{PGD-20 Accuracy [\%]} \\ \hline
    Robust Pretraining - Selfie & 79 & 6  \\ \hline
    Robust Pretraining - Rotation & \textcolor{blue}{\textbf{87}} & 18  \\ \hline
    Robust Pretraining - Jigsaw & 80 & 3  \\ \hline
    FlowGMM & 68 & 33  \\ \hline
     \hline
    Linear Head b=8 t=90 & 64 & 35  \\ \hline
    Linear Head b=8 t=30 & 72 & \textbf{49}  \\ \hline
    Linear Head b=7 t=10 & 82 & 5  \\ \hline
    Attention Head b=8 t=90 & 73 & \textcolor{blue}{\textbf{39}}  \\ \hline
    Attention Head b=8 t=30 & 85 & 25  \\ \hline
    Attention Head b=8 t=10 & \textbf{88} & 2  \\ \hline
    \end{tabular}
    }
\label{tab:CIFAR-10_Accuracies}
\end{table}

\begin{table}[t]
\caption{CIFAR-10 Classification under Adversarial Attacks. 
For each attack column, the best performing model's accuracy is marked in bold and the second best in blue.}
\centering
\resizebox{\linewidth}{!}{
\begin{tabular}{|c|c|c|c|c|c|c|c|c|c|}
\hline
\textbf{Model} & \textbf{Clean [\%]} & \textbf{FGSM [\%]} & \textbf{BIM [\%]} & \textbf{PGD-10 [\%]} & \textbf{PGD-20 [\%]} & \textbf{CW [\%]} & \textbf{FAB [\%]} & \textbf{APGD [\%]} & \textbf{AA [\%]} \\ \hline
ViT-B/16 (CIFAR-10 finetuned) & \textbf{97.88} & 44.01 & 0.76 & 52.04 & 0.27 & 18.98 & 0.01 & 0.00 & 0.00 \\ \hline
Linear Head b=8 t=30 & 77.00 & \textcolor{blue}{64.94} & \textcolor{blue}{53.16} & 77.35 & \textcolor{blue}{49.19} & 37.02 & 74.08 & 40.93 & 5.00 \\ \hline
Linear Head b=8 t=10 & 81.00 & \textbf{72.38} & \textbf{63.14} & \textcolor{blue}{81.27} & \textbf{60.64} & 23.92 & \textcolor{blue}{77.03} & \textbf{59.55} & 4.00 \\ \hline
Linear Head b=7 t=10 & \textcolor{blue}{82.00} & 39.10 & 8.99 & \textbf{81.59} & 5.59 & 51.00 & 76.81 & 6.92 & 5.00 \\ \hline
Attention Head b=8 t=50 & 81.00 & 47.52 & 31.18 & 80.96 & 34.05 & \textbf{77.41} & \textbf{78.75} & 33.40 & 46.00 \\ \hline
Attention Head b=8 t=90 & 73.00 & 47.98 & 36.94 & 73.79 & 39.43 & \textcolor{blue}{72.38} & 72.77 & \textcolor{blue}{44.38} & \textbf{56.00} \\ \hline
Attention Head b=7 t=90 & 71.00 & 44.13 & 33.79 & 71.06 & 34.81 & 69.80 & 69.84 & 41.09 & \textcolor{blue}{53.00} \\ \hline
\end{tabular}
}
\label{tab:CIFAR10_Attacks_Top3}
\end{table}

\begin{table}[t]
\caption{PASCAL VOC 2007 Object Detection Results of Diffusion-Based Models
}
\centering
\resizebox{\linewidth}{!}{
    \begin{tabular}{|c|c|c|c|c|c|}
    \hline
    \textbf{Head Type} & \textbf{Timestep} & \textbf{Clean mAP} & \textbf{CLS mAP} & \textbf{LOC mAP} & \textbf{CWA mAP}\\ \hline
    Convolution    & 60 & 59.24 & 36.46  & 43.59 & 39.85 \\ \hline
    Convolution    & 90 & 55.37 & 28.48  & 35.91 & 35.14 \\ \hline
    Convolution    & 120 & 45.43 & 41.85  & 42.11 & 44.03 \\ \hline
    Attention & 60 & 59.32 & 33.66 & 36.40 &  33.22 \\ \hline
    Attention & 90 & 58.47 & 28.10 & 30.63 &  25.62 \\ \hline
    Attention & 120 & 45.43 & 41.85 & 42.11 &  44.03 \\ \hline
    \end{tabular}
}
\label{tab:detection_results}
\end{table}

\begin{table}[t]
\caption{PASCAL VOC 2007 Object Detection Results of Adversarially Trained Diffusion-Based Models
}
\centering
\resizebox{\linewidth}{!}{
    \begin{tabular}{|c|c|c|c|c|c|}
    \hline
    \textbf{Head Type} & \textbf{Timestep} & \textbf{Clean mAP} & \textbf{CLS mAP} & \textbf{LOC mAP} & \textbf{CWA mAP}\\ \hline
    Convolution  & 60 & 40.15 & 33.51  & 31.45 & 34.49 \\ \hline
    Attention & 90 & 57.02 & 57.19 & 56.84 & 56.34 \\ \hline

    \end{tabular}
}
\label{tab:detection_adversarial}
\end{table}

\subsection{Ablation studies}
\label{sec:ablations}
To better understand the robustness of diffusion pretrainers, we perform additional experiments examining the effect of layer (block) and timestep choices on classification performance, as shown in \ref{fig2}.  

For the Linear Head, accuracy generally increases with block number, reaches a peak, and then decreases. The optimal block varies across attacks, but most attacks achieve their highest accuracy around blocks 7 and 8. In comparison, the Attention Head exhibits more uniform performance across blocks, showing less sensitivity to block selection, with the best accuracies consistently occurring around block 8 for all attacks.  

The effect of the diffusion timestep $t$ reveals two distinct behaviors across attacks for both heads. For stronger attacks, the model is more fragile at low noise levels (small $t$), with low accuracies initially. As $t$ increases, accuracy rises, reaching a peak (around timestep 150 for most attacks), and then declines. For weaker attacks, accuracy tends to decrease steadily as the timestep increases.

The choice of timestep entails a tradeoff: configurations that maximize clean accuracy often reduce robustness to certain attacks, and vice versa. Thus, the optimal block and timestep depend on the application requirements and whether priority is placed on clean accuracy or adversarial robustness. Selecting the operating point using AutoAttack is effective for prioritizing robustness, since it integrates multiple strong attacks and offers a balanced measure of robustness.

\section{CONCLUSION}
In this work, we studied the robustness of diffusion pretrainers, where a diffusion model is trained in an unsupervised manner on a  dataset, and then a classification or detection head is trained on top of feature maps extracted from the diffusion denoising network. We evaluated the robustness of such models under a wide range of strong adversarial attacks and showed that they offer out-of-the-box robustness without adversarial training. This robustness comes at very low computational cost, since only a small head is trained, making the approach particularly appealing for resource-constrained settings. For classification, we also examined the effect of layer and diffusion step choices on the results, finding that the timestep selection is especially. Overall, diffusion-based features offer an efficient and practical approach that delivers strong robustness gains with minimal overhead while maintaining competitive accuracy. We believe that our approach motivates future works to consider diffusion training as a self-supervised pretraining phase, which leads to robust models for the downstream tasks, thus, producing a  more robust system against adversarial attacks.

\pagebreak

\bibliographystyle{IEEEtran}
\bibliography{IEEEabrv,refs}

\end{document}